\documentclass[]{pasj01}
\usepackage{color}
\begin{document} 
\Received{}
\Accepted{}

\newcommand{\tr}{\textcolor{red}}

\title{Expected constraints on models of the epoch of reionization with the variance and skewness in redshifted 21cm-line fluctuations}

\author{Kenji \textsc{Kubota}\altaffilmark{1}%
\thanks{Example: Present Address is xxxxxxxxxx}}
\altaffiltext{1}{Department of Physics, Kumamoto University, Kumamoto, Japan}
\email{157d8005@st.kumamoto-u.ac.jp}

\author{Shintaro \textsc{Yoshiura}\altaffilmark{1}}
\altaffiltext{2}{Department of Physics, Nagoya University, Aichi, Japan}

\author{Hayato \textsc{Shimabukuro}\altaffilmark{1,2}}

\author{Keitaro \textsc{Takahashi}\altaffilmark{1}}


\maketitle

\begin{abstract}
Redshifted 21cm-line signal from neutral hydrogens in the intergalactic medium (IGM) gives a direct probe of the epoch of reionization (EoR). In this paper, we investigate the potential of the variance and skewness of the probability distribution function of the 21cm brightness temperature for constraining EoR models. These statistical quantities are simple, easy to calculate from the observed visibility and thus suitable for the early exploration of the EoR with ongoing telescopes such as the Murchison Widefield Array (MWA) and LOw Frequency ARray (LOFAR). We show, by performing Fisher analysis, that the variance and skewness at $z=7-9$ are complementary to each other to constrain the EoR model parameters such as the minimum virial temperature of halos which host luminous objects, ionizing efficiency and mean free path of ionizing photons in the IGM. Quantitatively, the constraining power highly depends on the quality of the foreground subtraction and calibration. We give a best case estimate of the constraints on the parameters, neglecting the systematics other than the thermal noise.
\end{abstract}

\section{Introduction}

After recombination, a large amount of neutral hydrogen in intergalactic medium (IGM) was reionized by stars and galaxies which emit UV and X-ray photons. This important phase of the universe, the Epoch of Reionization (EoR), has attracted much attention in a broad community of astrophysics and cosmology \citep{2006PhR...433..181F,2012RPPh...75h6901P}. The analysis of the Gunn-Peterson effect \citep{1965ApJ...142.1633G} in the spectra of high-z quasars indicates that the reionization of hydrogen was completed by $z \approx 6$ \citep{2006AJ....132..117F}. On the other hand, the integrated Thomson scattering optical depth of the CMB photons implies the instantaneous reionization redshift of $z \sim 8.8$ \citep{2015arXiv150201589P}. Contrastingly, we have poor information on the start and the progress of the reionization and the nature of ionizing sources.

A direct observation of neutral hydrogen of the IGM with the 21cm line is expected to be a powerful tool to probe the cosmic dawn and the reionization. There are several ongoing telescopes which are beginning observations and getting constraints on the fluctuations in the 21cm signal: the MWA \citep{2009IEEEP..97.1497L,2013PASA...30....7T,2013MNRAS.429L...5B}, the LOFAR \citep{2013A&A...556A...2V,2013MNRAS.435..460J} and the Precision Array for Probing the Epoch of Reionization (PAPER) \citep{Jacobs2015,2015ApJ...809...61A}. Although their sensitivities are not enough to obtain images of 21cm signal during the EoR, the statistical information on the 21cm-signal fluctuations is expected to be obtained after a sufficient subtraction of the foreground emission and some upper bounds on the power spectrum have already been placed \citep{2015PhRvD..91l3011D}. Much higher sensitivity is required for the imaging of 21cm signal and the Square Kilometre Array (SKA) will be the ultimate telescope for this purpose \citep{2015aska.confE.171C}.

The power spectrum has often been used to analyze the statistical properties of the fluctuations in 21cm signal \citep{2006PhR...433..181F,2007MNRAS.376.1680P,2008ApJ...689....1S,2010A&A...523A...4B,2013MNRAS.431..621M,2014ApJ...782...66P,2006ApJ...653..815M,2010MNRAS.405.2492H,2015MNRAS.449.4246G}. The variance of the probability distribution function (PDF) of the fluctuations is also a standard quantity which is simple to compute both theoretically and observationally, because it can be calculated by an integration of the power spectrum with respect to the wave number \citep{2014MNRAS.443.1113P}. Both of them were shown to be very useful tools to probe the global history of the reionization and to constrain the parameters of EoR models.

The bispectrum and the skewness of the PDF, which is an integral of the bispectrum with respect to the wave number, are also fundamental statistical quantities which characterize the fluctuations \citep{2015arXiv150701335S}. These can measure the non-Gaussianity of the fluctuations, which are naturally generated in the highly-nonlinear processes of the reionization and cannot be captured by the power spectrum and variance. In \citet{2015MNRAS.451..266Y}, the thermal noise for the bispectrum observation was estimated and it was shown that the above ongoing telescopes have enough sensitivity to detect the bispectrum at large scales. Further, \citet{2015MNRAS.451..467S} showed that the skewness is a good indicator of the onset of the X-ray heating of the IGM and the nature of the ionizing sources \citep{2007MNRAS.379.1647W,2014MNRAS.443.3090W,2015MNRAS.454.1416W,2015MNRAS.449.3202W}. The 21cm PDF \citep{2010MNRAS.406.2521I} and the 21cm difference PDF \citep{2012arXiv1209.5751P,2008MNRAS.384.1069B} are also shown to be useful tools to study the reionization scenario.

In this paper, we investigate, using Fisher analysis, the potential of the variance and skewness to constrain some of the key parameters of an EoR model. Both of these are simple and suitable for the early exploration of the EoR with 21cm signal. We generate maps of the brightness temperature using a public semi-analytic code 21cmFAST \citep{2011MNRAS.411..955M} and estimate expected constraints on the model parameters by future observation data with the MWA and LOFAR. As explained above, the variance and skewness have complimentary information so that the combination will give us effective constraints on the parameters.

The paper is organized as follows: in section 2, we introduce the fluctuations in the brightness temperature and their statistical characterization. We explain our methodology in section 3. First, we describe the simulation code we have used and show the basic properties of the variance and skewness. Then, we describe the thermal noise to these quantities and explain Fisher analysis for the estimation of the parameter constraints. The results are shown in section 4. Finally, we summarize our results and give discussion in section 5.

\section{Statistical characterization of 21cm line fluctuations}

The observable quantity of the redshifted 21cm line is brightness temperature $\delta T_b$ which is defined as the contrast between the spin temperature $T_S$ and the background CMB temperature $T_\gamma$ \citep{2006PhR...433..181F}:
\begin{eqnarray}
\delta T_b(z)&=&\frac{T_S-T_{\gamma}}{1+z}(1-\mathrm{e}^{{-\tau_{\nu_0}}}) \nonumber \\
                &\approx&27x_{\rm HI}(1+\delta_{\rm m})\Bigl(\frac{H}{dv_r /dr+H}\Bigr)\Bigl(1-\frac{T_{\gamma}}{T_S}\Bigr) \nonumber \\ &&\times\Bigl(\frac{1+z}{10}\frac{0.15}{\Omega_Mh^2}\Bigr)^{\frac{1}{2}}\Bigl(\frac{\Omega_bh^2}{0.023}\Bigr)\ [\rm mK],
\end{eqnarray}
where $x_{\rm HI}$ is the neutral hydrogen fraction, $\delta_{\rm m}$ is the matter density fluctuation, $H$ is the Hubble parameter and $dv_r/dr$ is the gradient of the proper velocity along the line of sight. The spin temperature is determined by the of number-density ratio of the singlet and triplet of a neutral hydrogen atom:
\begin{equation}
\frac{n_1}{n_0}=\frac{g_1}{g_0}\exp\Bigl(-\frac{h \nu_{21}}{k T_S}\Bigr),
\end{equation}
where $h$ is the Planck constant, $k$ is the Boltzmann constant, $n_0$, $n_1$ are the number density and $g_0$, $g_1$ are the statistical weight of singlet and triplet, respectively. In order to analyze $\delta T_b$ statistically we define spacial fluctuation:
\begin{equation}
\delta_{21}({\bf x},z)\equiv\frac{\delta T_b({\bf x},z)}{\overline{\delta T_b}(z)}-1,
\end{equation}
where $\overline{\delta T_b}(z)$ is the spatial average of $\delta T_b$. We introduce variance and skewness in the next two subsections.

\subsection{power spectrum and variance}

The power spectrum $P(\bf k)$ is defined as:
\begin{eqnarray}
\langle\tilde{\delta}_{21}({\bf k_1})\tilde{\delta}_{21}({\bf k_2})\rangle=(2\pi)^3\delta_D({\bf k_1+k_2})P({\bf k_1}),
\end{eqnarray}
where $\langle\cdots\rangle$ represents the ensemble average, $\delta_D(\bf k)$ is Dirac's delta function, and the $\tilde{\delta}_{21}({\bf k})$ is the Fourier transform of $\delta_{21}({\bf x},z)$. 

On the other hand, the variance is defined as,
\begin{equation}
\sigma^2 =
\frac{1}{N}\sum_{i=1}^N (\delta T_{b,i}-\overline {\delta T_b})^2,
\label{eq:variance1}
\end{equation}
where $N$ is the number of pixels. Furthermore, the variance can be calculated by an integral of power spectrum with respect to wave number,
\begin{equation}
\sigma^2 =
(\overline{\delta T_b})^2 \int \frac{d^3 k}{(2\pi)^3}P({\bf k}),
\label{eq:variance2}
\end{equation}
where the integration range is determined by the angular resolution and observation area. Although the above two expressions are mathematically equivalent, Eq. (\ref{eq:variance2}) is more practical for observations with interferometer because the Fourier components, $\tilde{\delta}_{21}({\bf k})$, can directly be obtained from visibility without aperture synthesis.

\subsection{bispectrum and skewness}

The bispectrum $B(\bf k)$ is defined as:
\begin{eqnarray}
\langle
\tilde{\delta}_{21}({\bf k_1})
\tilde{\delta}_{21}({\bf k_2})
\tilde{\delta}_{21}({\bf k_3})
\rangle
= (2\pi)^3 \delta_D({\bf k_1 + k_2 + k_3}) B({\bf k_1,k_2}).
\end{eqnarray}
The delta function guarantees the triangle condition, $\bf k_1 + k_2 + k_3 = 0$. In literature, bispectra for specific shapes of triangles are often considered, such as equilateral type ($k_1 = k_2 = k_3$), folded type ($k_1 + k_2 = k_3$) and squeezed type ($k_1, k_2 \gg k_3$).

Skewness $\gamma$ is defined as,
\begin{equation}
\gamma=\frac{1}{N}\sum_{i=1}^N (\delta T_{b,i}-\overline {\delta T_b})^3.
\label{eq:skewness1}
\end{equation}
Similar to the case of variance, skewness can be calculated from an integral of the bispectrum as,
\begin{equation}
\gamma=(\overline{\delta T_b})^3 \int \frac{d^3 k_1}{(2 \pi)^3}\int \frac{d^3 k_2}{(2 \pi)^3} B({\bf{k_1}, \bf{k_2}, - \bf{k_1} - \bf{k_2}}).
\label{eq:skewness2}
\end{equation}
Here it should be noted that the integration contains all types of the bispectrum.

While skewness is a simple statistical quantity as variance, a lot of information is lost through the process of integration. Thus, in order to increase information and still keep the simpleness, we consider two kinds of "skewness", $\gamma_{\rm e}$ and $\gamma_{\rm f}$, which counts contributions only from equilateral- and folded-type bispectra, respectively.

\section{Method}

\subsection{simulation}

We generate $\delta T_b$ map at each redshift by a semi-analytic simulation code called 21cmFAST \citep{2007ApJ...669..663M,2011MNRAS.411..955M}. In this code, the matter density is initialized with $1800^3$ dark matter particles at $z=300$ and the density and velocity fields are evolved through Zel'dovich approximation. Then, the excursion set formalism is used to count dark matter haloes in the density field and luminous objects are assumed to reside in each halo. Considering the emission of UV, ionizing and X-ray photons, thermal and ionizing states of intergalactic medium (IGM) are calculated and the maps of brightness temperature and other physical quantities are provided. The results are in good agreement with hydrodynamical simulations on scales larger than 1 Mpc \citep{2011MNRAS.411..955M}.

There are three phenomenological parameters which significantly affect the structure and evolution of IGM.
\begin{itemize}
\item $T_{\rm vir, min}$: the minimum virial temperature which gives the minimum mass of halos which host luminous objects. Physically, this is determined by the cooling process of gas and significantly affected by the feedback from star formation. This quantity effectively parametrizes the efficiency of the feedback.
\item $\zeta_{\rm ion}$: ionizing efficiency of luminous objects, which includes the number of emitted ionizing photons per baryon and the escape fraction from galaxies.
\item $R_{\rm mfp}$: the ionizing photon horizon which can be chosen to match the extrapolated ionizing photon mean free path in the ionized IGM. This represents the maximum distance which ionizing photons are allowed to propagate and effectively parametrizes the clumpiness of the IGM and determines the maximum size of ionized bubbles.
\end{itemize}
We form the evolved simulation boxes of $600~{\rm cMpc}^3$ with $600^3$ grids. We focus on the redshift range of $7 \leq z \leq 9$ to which ongoing telescopes are most sensitive. The fluctuations in $T_S$ are neglected assuming $T_S \gg T_\gamma$ at these redshifts \citep{2012RPPh...75h6901P}. The variance and skewness of $\delta T_b$ are computed by integrating the power spectrum and bispectrum, respectively, which are obtained from the Fourier transform of $\delta T_b$ map. In the Fisher analysis given below, we set the fiducial values of these parameters as $T_{\rm vir, min} = 10^4{\rm K}$, $\zeta_{\rm ion} = 15.0$ and $R_{\rm mfp} = 30{\rm Mpc}$. For these parameter values, the reionization is completed at $z \sim 6$.

\begin{figure}[t]
\begin{center}
\includegraphics[width=8cm]{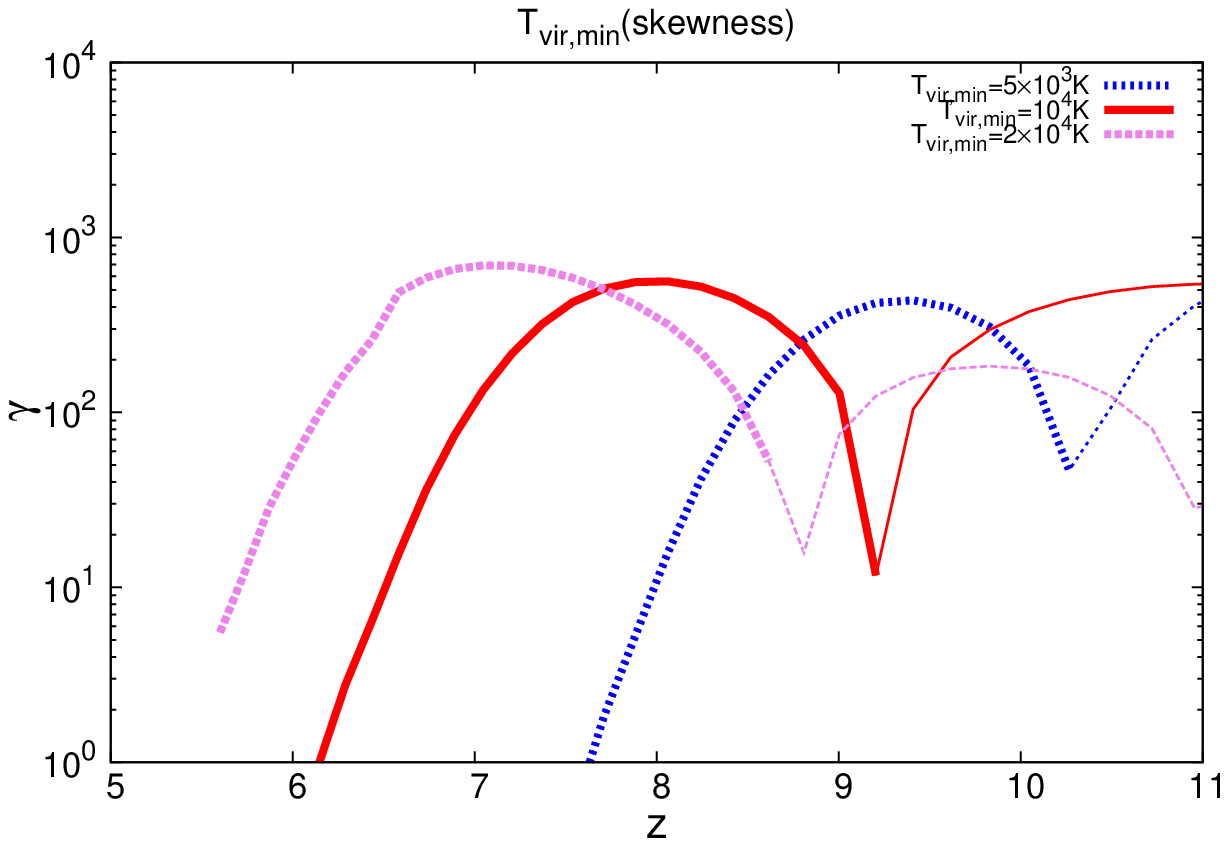}
\includegraphics[width=8cm]{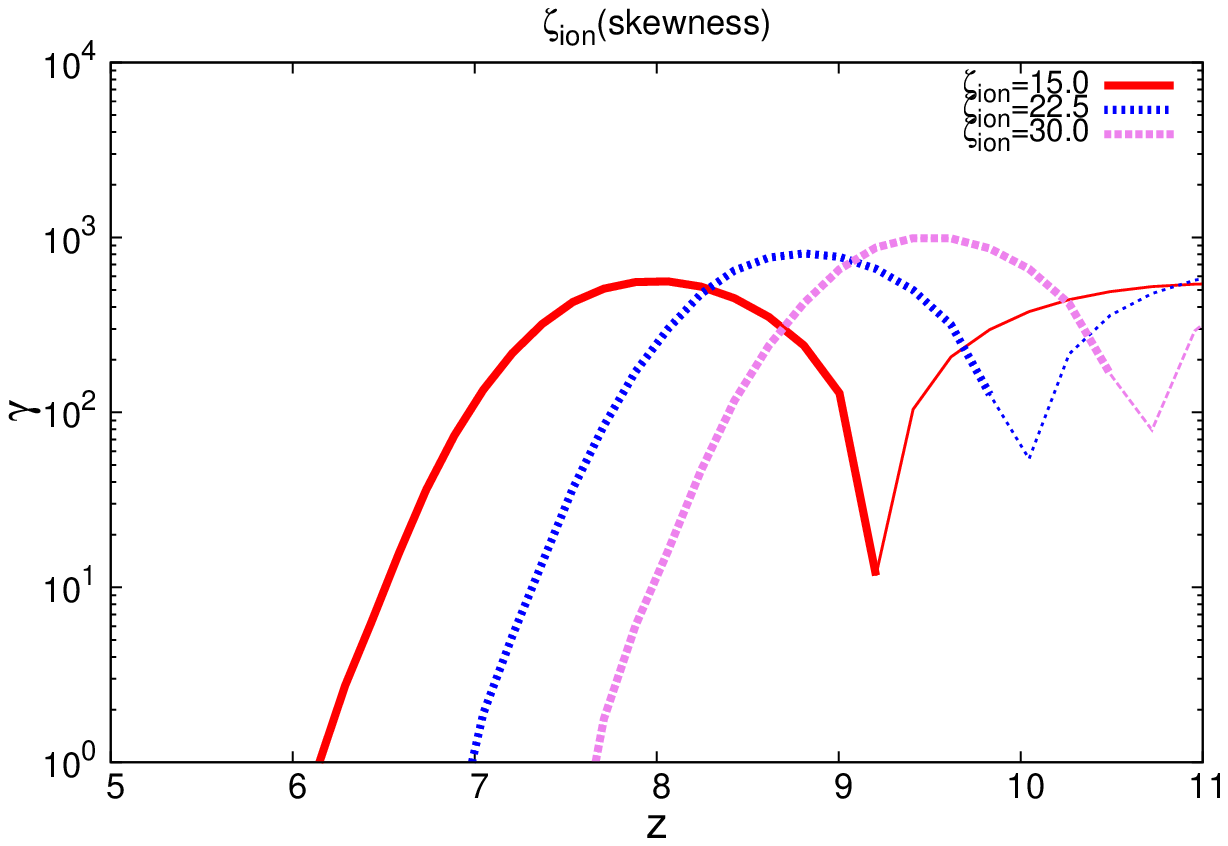}
\includegraphics[width=8cm]{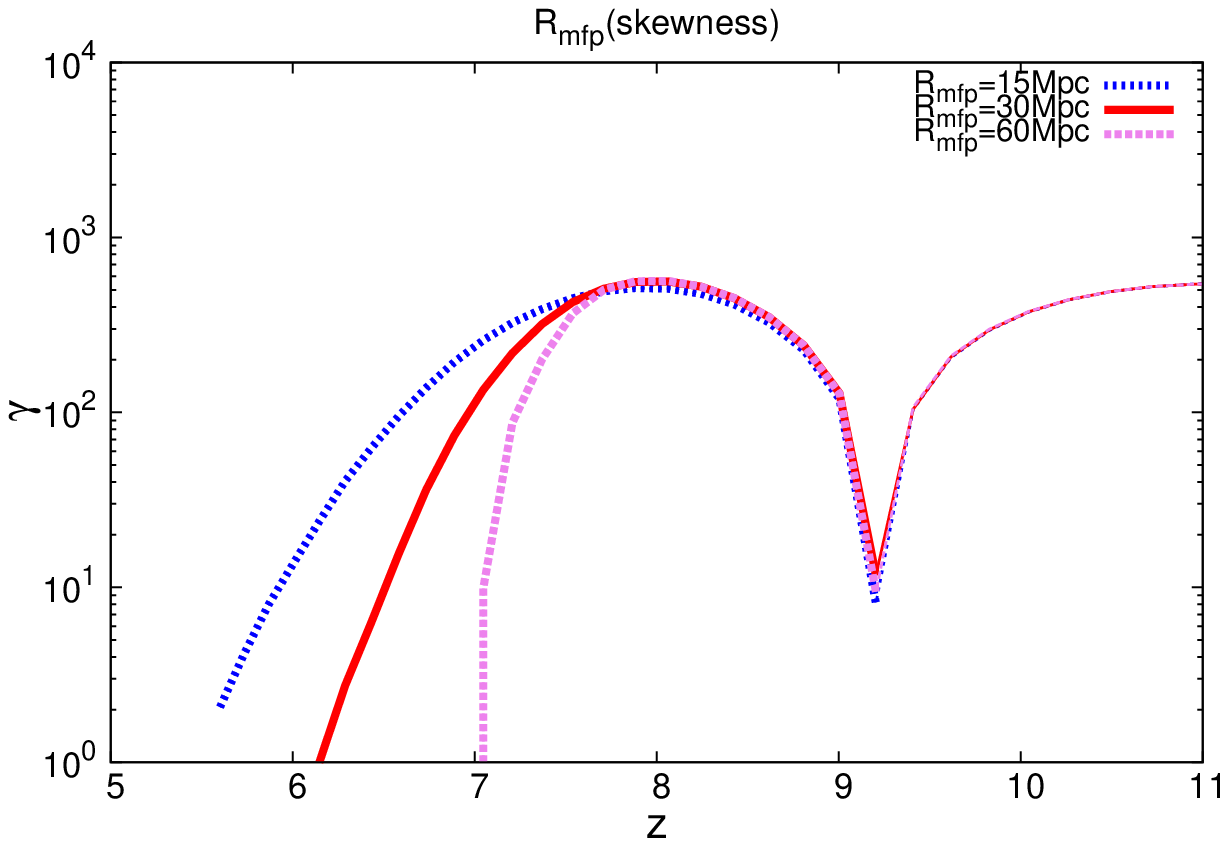}
\end{center}
\vspace{10mm}
\caption{Evolution of the skewness $\gamma$ and its dependence of the above parameters: $T_{\rm vir, min}$ (top), $\zeta_{\rm ion}$ (mid) and $R_{\rm mfp}$ (bottom). The thick and thin lines represent the positive and negative values, respectively.}
\label{fig:skew-evo1}
\end{figure}

Fig. \ref{fig:skew-evo1} shows the evolution of the skewness $\gamma$ of brightness temperature and its dependence of the above parameters as a function of redshift : $T_{\rm vir, min}$ (top), $\zeta_{\rm ion}$ (mid) and $R_{\rm mfp}$ (bottom). In each figure, the solid line represents the fiducial model. During the EoR we are focusing on ($z \lesssim 10$), the fluctuations in brightness temperature are mostly contributed from those in neutral hydrogen fraction. The skewness increases as reionization proceeds and reaches its peak at $z \sim 8$ where about half of hydrogen is ionized. After that, the skewness decreases and approaches to zero as the entire IGM is eventually ionized at $z \sim 6$.

For a smaller value of $T_{\rm vir,min}$, the number of halos with luminous objects increases so that reionization proceeds more quickly and the peak shifts to a higher redshift. Increasing $\zeta_{\rm ion}$ has a similar effect. As for $R_{\rm mfp}$, it does not affect the fluctuations until $z \sim 8$. This is because ionizing photons emitted from galaxies ionize the local neutral hydrogen in the early phase of reionization. In the later phase, larger ionized bubbles are formed for a larger value of $R_{\rm mfp}$ and, therefore, reionization is completed earlier.

\begin{figure}[t]
\begin{center}
\includegraphics[width=8cm]{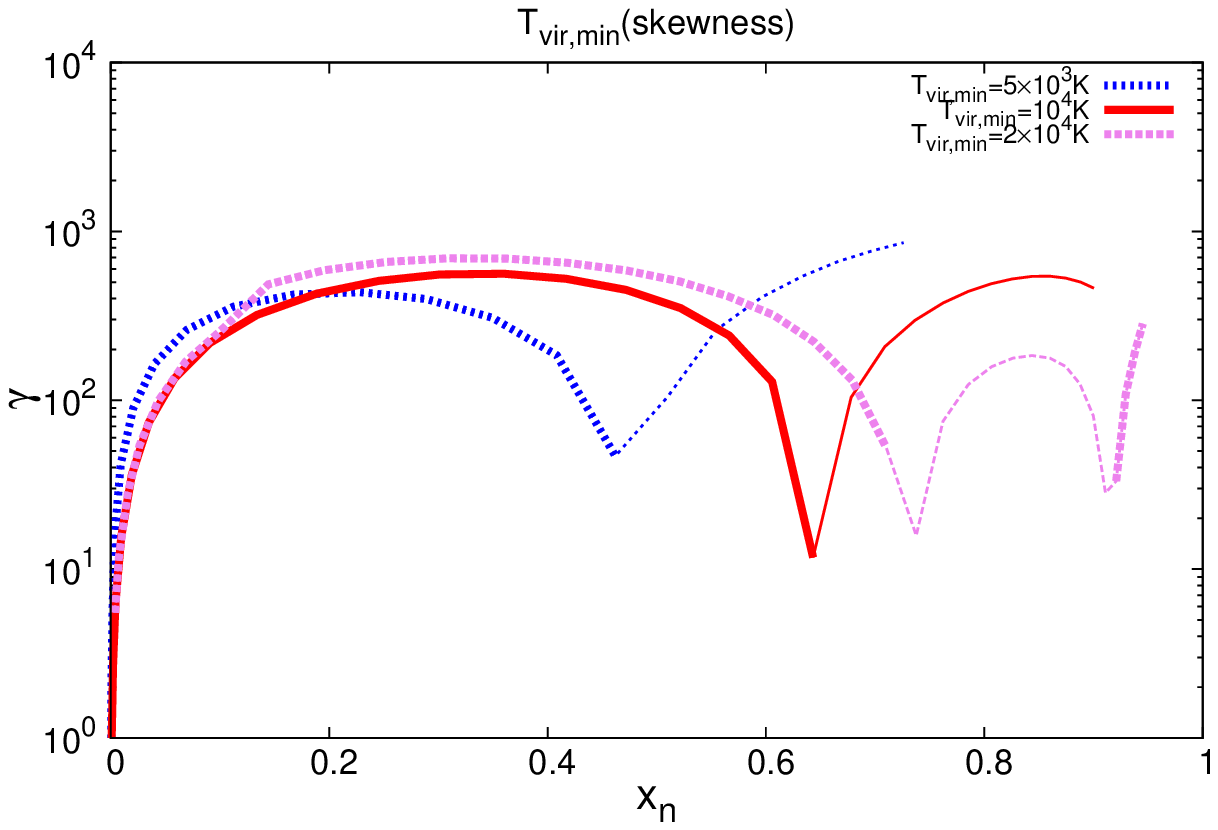}
\includegraphics[width=8cm]{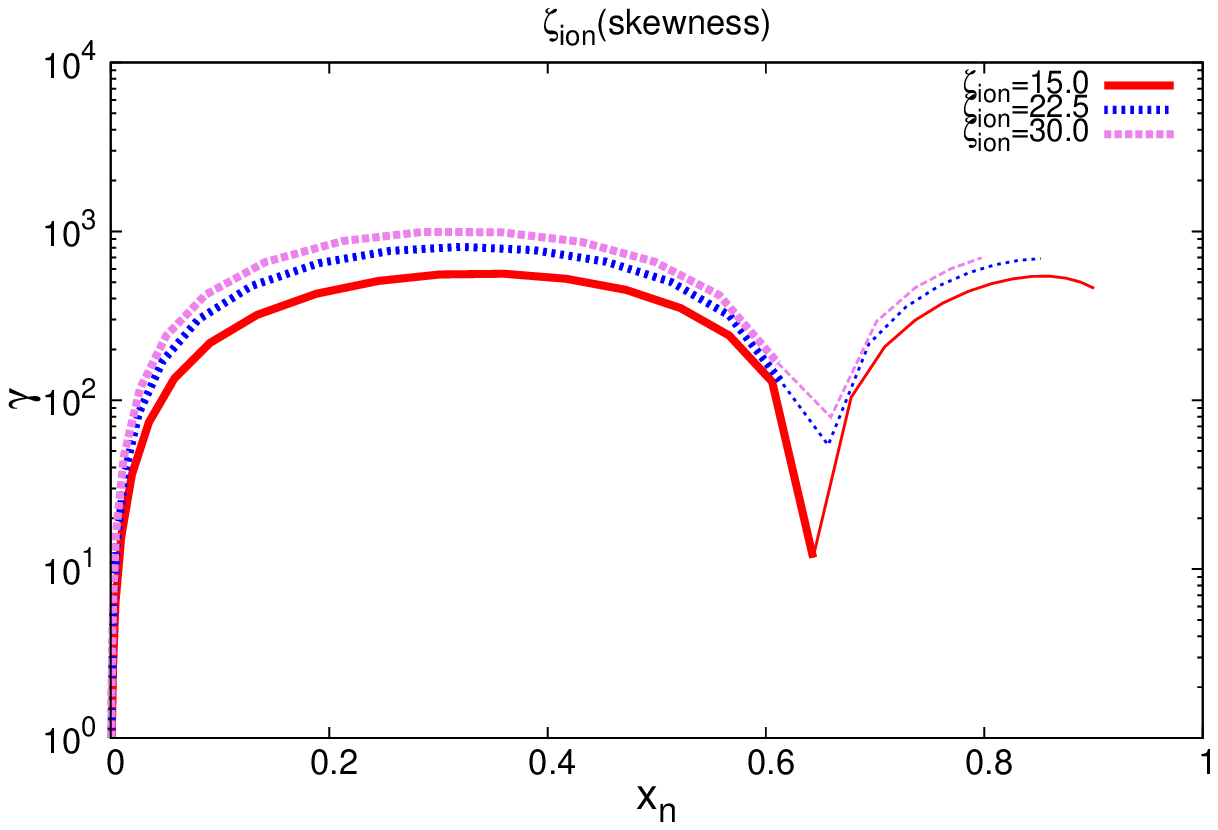}
\includegraphics[width=8cm]{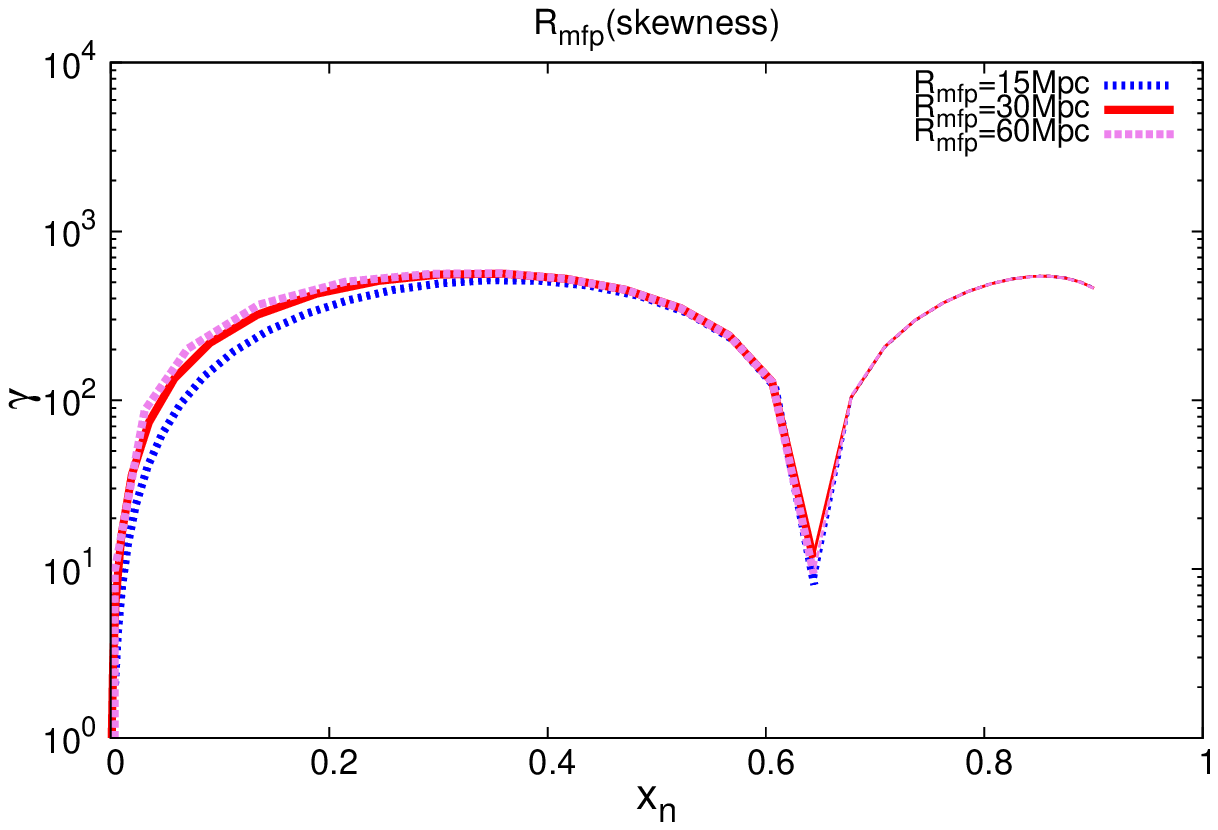}
\end{center}
\vspace{10mm}
\caption{Same as Fig. \ref{fig:skew-evo1} but as a function of the neutral fraction.}
\label{fig:skew-evo2}
\end{figure}

Similarly, Fig. \ref{fig:skew-evo2} shows the evolution of the skewness $\gamma$ as a function of the neutral fraction. We find that the dependence of the skewness on the model parameters are generally weaker compared with that in Fig. \ref{fig:skew-evo1}. Especially, the dependence on $R_{\rm mfp}$ almost disappears, which indicates that $R_{\rm mfp}$ affects the skewness only through the change in the ionizing history, rather than the properties of the fluctuations. On the other hand, $\zeta_{\rm ion}$ does not change the overall shape and affects only the normalization, while $T_{\rm vir,min}$ changes significantly the behavior of the skewness in the early phase of reionization.

\subsection{thermal noise in variance and skewness}

We need observation errors to perform Fisher analysis. In this paper, we take only thermal noise into account because other factors such as the effect of foreground and incomplete calibration for skewness (bispectrum) are yet to be studied. These systematics are currently dominant over the thermal noise and, in this sense, our estimation below gives the least errors.

The thermal noise for the variance and skewness can be calculated from those for the power spectrum \citep{2006ApJ...653..815M} and bispectrum \citep{2015MNRAS.451..266Y}, respectively, as,
\begin{eqnarray}
\sigma^2_{\rm v}
&=& \sum_{i=k_{\rm min}}^{k_{\rm max}}
    \left( \frac{k^3}{2 \pi^2} P_{\rm N}(k) d\log k \right)^2, \\
\sigma^2_{\rm s}
&=& \sum_{i=k_{\rm min}}^{k_{\rm max}}
    \left( \frac{0.1 k^6}{4 \pi^4} B_{\rm N}(k) d\log k \right)^2,
\end{eqnarray}
where, $P_{\rm N}(k)$ and $B_{\rm N}(k)$ is the noise power spectrum and bispectrum, respectively. We set $k_1 = k_2 = k$ and $dk_2 = 0.1 k$. To calculate the noise, we assume that the MWA has 256\footnote{Currently, the MWA has 128 total tiles but it plans to double the number in a couple of years.} antennae within a radius of 750 m with $r^{-2}$ distribution \citep{Bowman2006} and that the LOFAR has 24 antennae within a radius of 2,000 m with $r^{-2}$ distribution \citep{2013A&A...556A...2V}. Further, we assume 1,000 hours for the total observing time and 6 MHz bandwidth.

\subsection{Fisher analysis}

We perform Fisher analysis to estimate constraints on parameters of 21cm FAST expected to be obtained in future observation data \citep{2009arXiv0906.4123C,2010LNP...800..147V}. Assuming that the likelihood function is Gaussian form, the Fisher matrix is defined as,
\begin{equation}
F_{ij}
= \frac{1}{2} \Bigl \langle \frac{\partial^2\chi^2}{\partial p_i\partial p_j}\Bigr \rangle \Bigl|_{\vec{p}=\vec{p}_{\rm fid}},
\end{equation}
where $\vec{p} = (T_{\rm vir, min}, \zeta_{\rm ion}, R_{\rm mfp})$ is the model parameter vector and $\vec{p}_{\rm fid}$ is the fiducial vector. For $N$ independent observation data $x_k(\vec{p}) ~ (k=1,\cdots,N)$, $\chi^2$ value is written as,
\begin{equation}
\chi^2(\vec{p})
= \sum_k^N \frac{[x_k(\vec{p}) - x_k(\vec{p}_{\rm fid})]^2}{\sigma_k^2},
\end{equation}
where $\sigma_k$ is the error in $x_k(\vec{p})$. Here, we consider, for observable quantities, the variance, the skewness from equilateral-type bispectrum and skewness from folded-type bispectrum at $z = 7, 8$ and $9$. Then, the Fisher matrix can be rewritten as,
\begin{equation}
F_{ij}
= \sum_k^N \frac{1}{\sigma_k^2}
           \frac{\partial x_k(\vec{p})}{\partial p_i}
           \frac{\partial x_k(\vec{p})}{\partial p_j}
           \Bigl|_{\vec{p}=\vec{p}_{\rm fid}}.
\end{equation}
Given the Fisher matrix, the covariance matrix is given by the inverse matrix, $C_{ij} = F_{ij}^{-1}$ and we can estimate the expected 1-$\sigma$ error of $p_{i}$ from $\sqrt{F_{ii}^{-1}}$.

\section{Results}

\begin{figure}[t]
\begin{center}
\includegraphics[width=8cm]{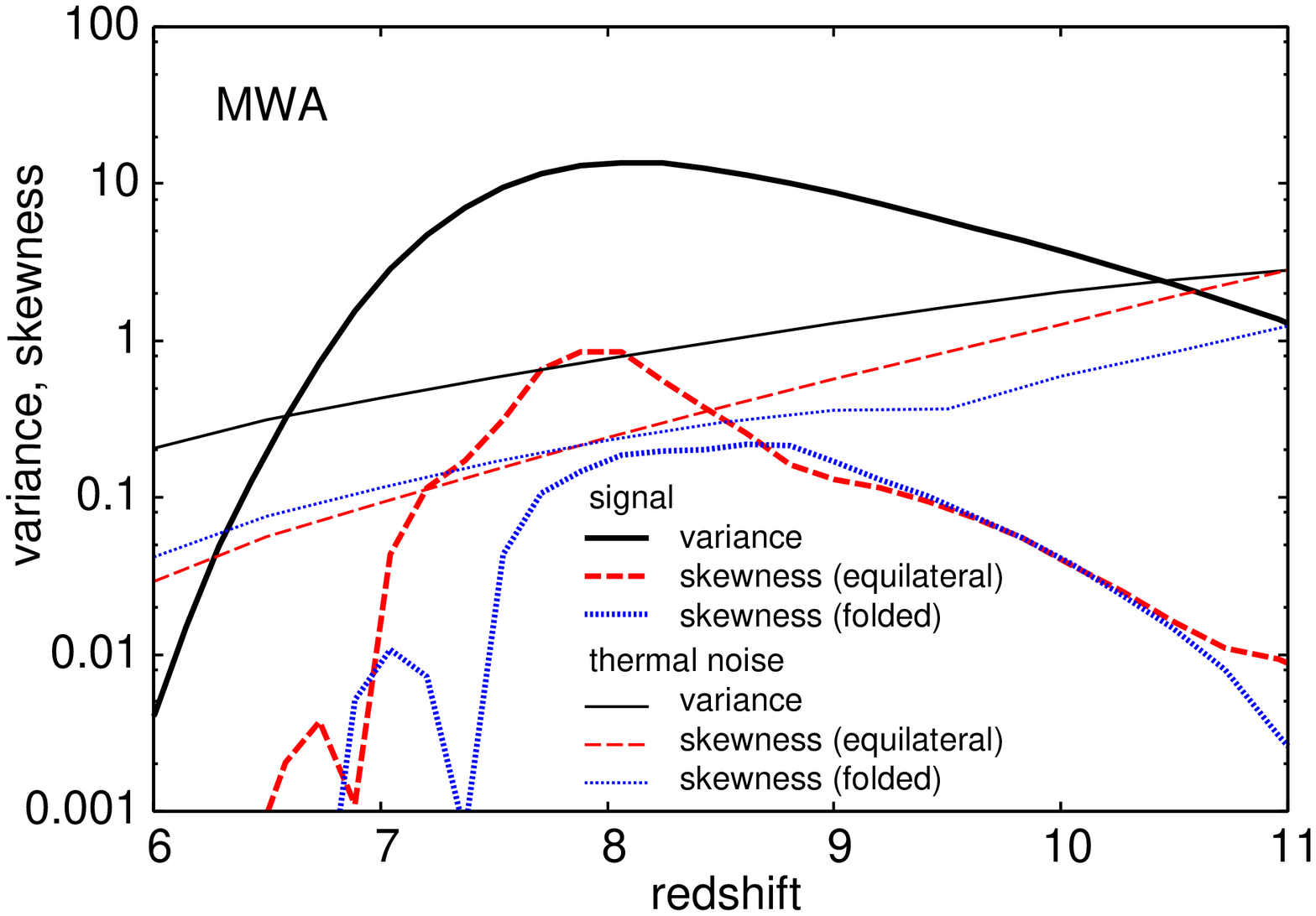}
\includegraphics[width=8cm]{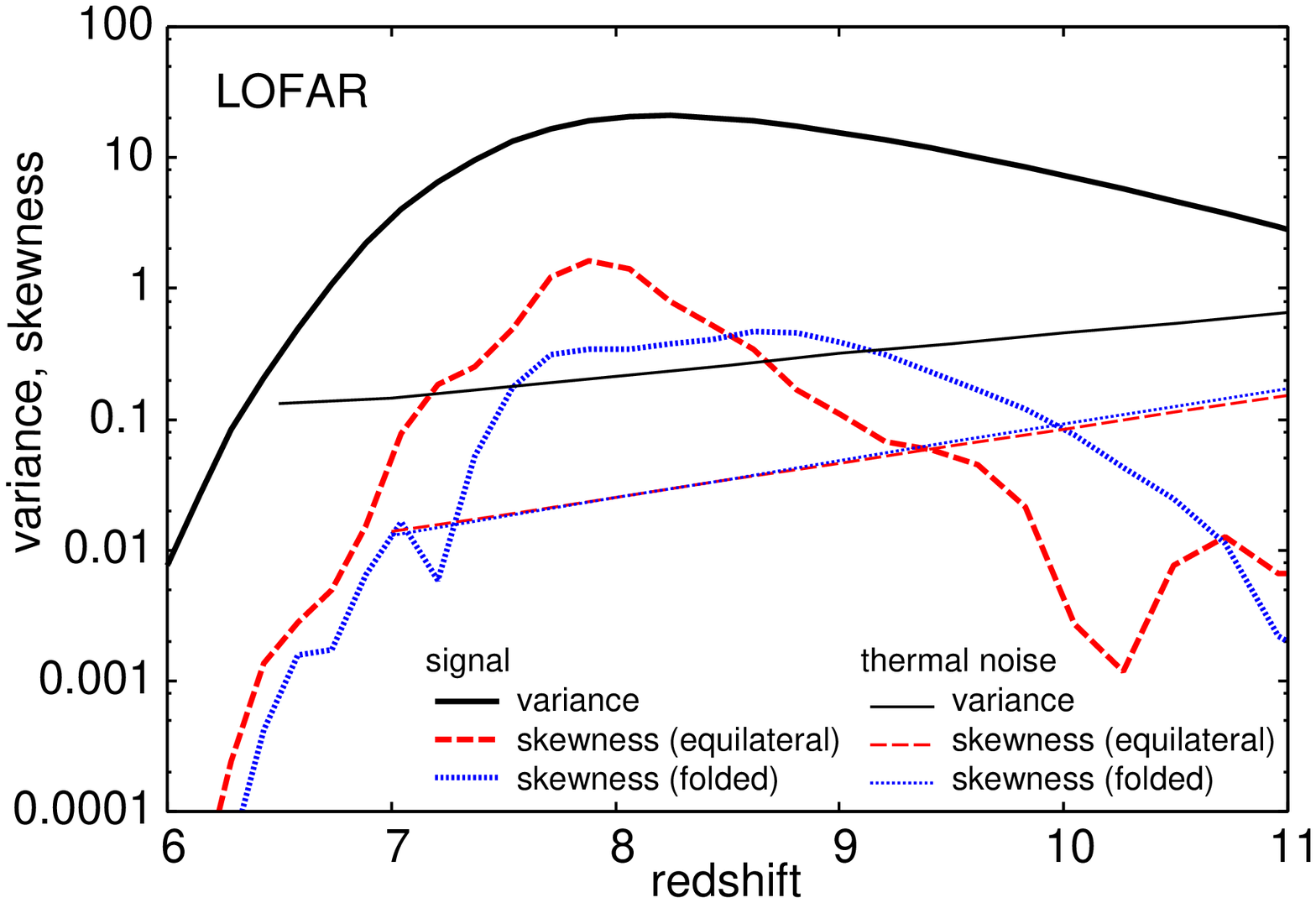}
\end{center}
\caption{Evolution of the variance, the equilateral-type and folded-type skewnesses for the MWA (top) and LOFAR (bottom) observations. The signal skewness is shown the absolute value.The thermal-noise curves are also shown. For the LOFAR, the noise curves for the two skewnesses are almost overlapping with each other.}
\label{fig:skew-evo3}
\end{figure}

In this section, we show the results of Fisher analysis. First of all, we argue the detectability of the variance and skewness with the MWA and LOFAR. It should be noted that the signal variance and skewness depend on the angular resolution and channel width of the telescope, and observation area due to the scale dependence of the power spectrum and bispectrum, respectively. This effect can be easily accounted in Eqs. (\ref{eq:variance2}) and (\ref{eq:skewness2}) by limiting the integration range according to the specification of the observation. Taking the array distribution and field-of-view (FoV) of the telescopes, we consider a deep observation of a $(4~{\rm deg})^2$ field which is much smaller than and comparable to the FoV of the MWA and LOFAR, respectively, and take $0.01 \lesssim k \lesssim 0.07~{\rm Mpc}^{-1}$ for the MWA and $0.03 \lesssim k \lesssim 0.1~{\rm Mpc}^{-1}$ for the LOFAR. 

Fig. \ref{fig:skew-evo3} shows the evolution of the variance, the equilateral-type and folded-type skewnesses for the MWA and LOFAR observations. The behavior of these quantities are slightly different from each other and the folded-type skewness is peaked at a higher redshift than the other quantities. The thermal-noise curves for the variance, the equilateral- and folded-type skewnesses are also shown. We can expect that the signal-to-noise (S/N) ratios are relatively large for a redshift range of $7 \lesssim z \lesssim 9$. Thus, we focus our analysis on this redshift range hereafter.

\begin{table}
\begin{center}
\caption{Signal-to-noise ratio of the variance, the skewness from equilateral-type bispectrum, $\gamma_e$, and the skewness from folded-type bispectrum, $\gamma_f$, for the MWA and LOFAR.}
\label{table:SN}
\vspace{0.5cm}
\begin{tabular}{|c|c|c|c|c|} \hline
      & $z$ & $\sigma^2$ & $\gamma_e$ & $\gamma_f$ \\ \hline \hline
      & 7   & 3.0        & 0.95       & 0.21       \\ \cline{2-5}
MWA   & 8   & 9.5        & 6.7        & 0.24       \\ \cline{2-5}
      & 9   & 3.2        & 0.70       & 0.63       \\ \hline
      & 7   & 16         & 8.2        & 0.98       \\ \cline{2-5}
LOFAR & 8   & 62         & 120        & 11         \\ \cline{2-5}
      & 9   & 26         & 18         & 23         \\ \hline
\end{tabular}
\end{center}
\end{table}

Based on Fig. \ref{fig:skew-evo3}, we show the S/N ratios of three observable quantities, $\sigma^2$, $\gamma_e$ and $\gamma_f$, at $z = 7, 8$ and $9$ for the fiducial model parameters in Table \ref{table:SN}. The S/N for LOFAR is much larger than that for the MWA due to its large effective area and better angular resolution. The MWA is expected to be able to detect $\sigma^2$ and $\gamma_e$, while the LOFAR will be able to detect all the three quantities.

\begin{figure}
\begin{center}
\includegraphics[width=10cm]{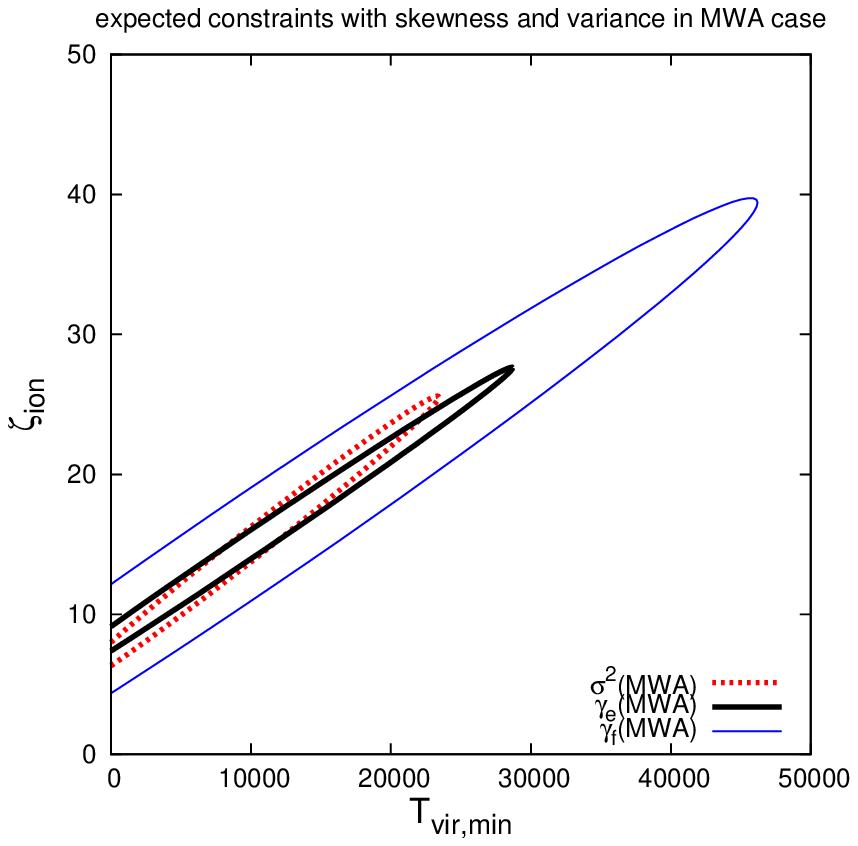}
\includegraphics[width=10cm]{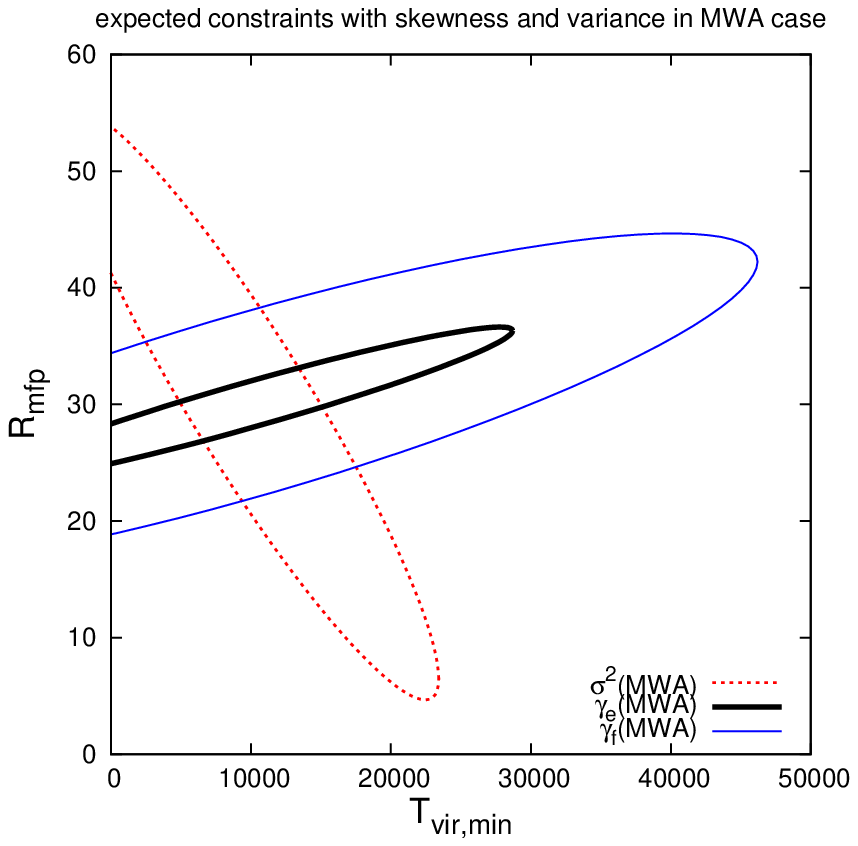}
\includegraphics[width=10cm]{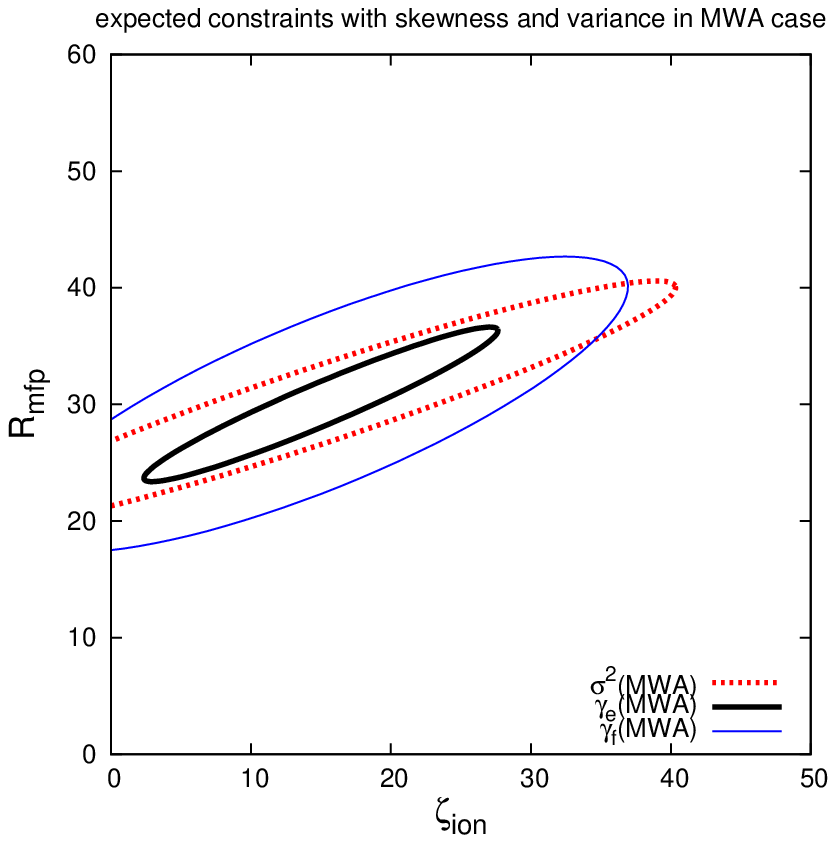}
\vspace{13mm}
\end{center}
\caption{Expected 1-$\sigma$ constraints on EoR parameters with the skewness and variance for 1000-hour observation of the MWA. The thick solid, thin solid and dotted lines represent the constraints from the equilateral-type skewness, the folded-type skewness and the variance, respectively.}
\label{fig:MWA1}
\end{figure}

Fig. \ref{fig:MWA1} represents the expected 1-$\sigma$ constraints on the three parameters, $T_{\rm vir, min}, \zeta_{\rm ion}, R_{\rm mfp}$, for MWA observation. Contributions from $\sigma^2$, $\gamma_e$ and $\gamma_f$ are separately plotted while a summation was taken in terms of the redshift. The equilateral- and folded-type skewnesses have the same degeneracy in all the panels, while the constraint from the former is much better than that from the latter due to the larger S/N ratio. The degeneracy is partially broken by the variance, especially in the $T_{\rm vir, min}$-$R_{\rm mfp}$ plane. The combination of the three quantities leads to the parameter constraints of $T_{\rm vir,min} = 10^4 \pm 6000~{\rm K}$, $\zeta_{\rm ion} = 15.0 \pm 13.0$ and $R_{\rm mfp} = 30.0 \pm 3.0~{\rm Mpc}$. The value of $R_{\rm mfp}$, equivalently the clumpiness of the IGM, is well constrained, although the constraint on the other two are relatively poor. 

\begin{figure}
\begin{center}
\includegraphics[width=10cm]{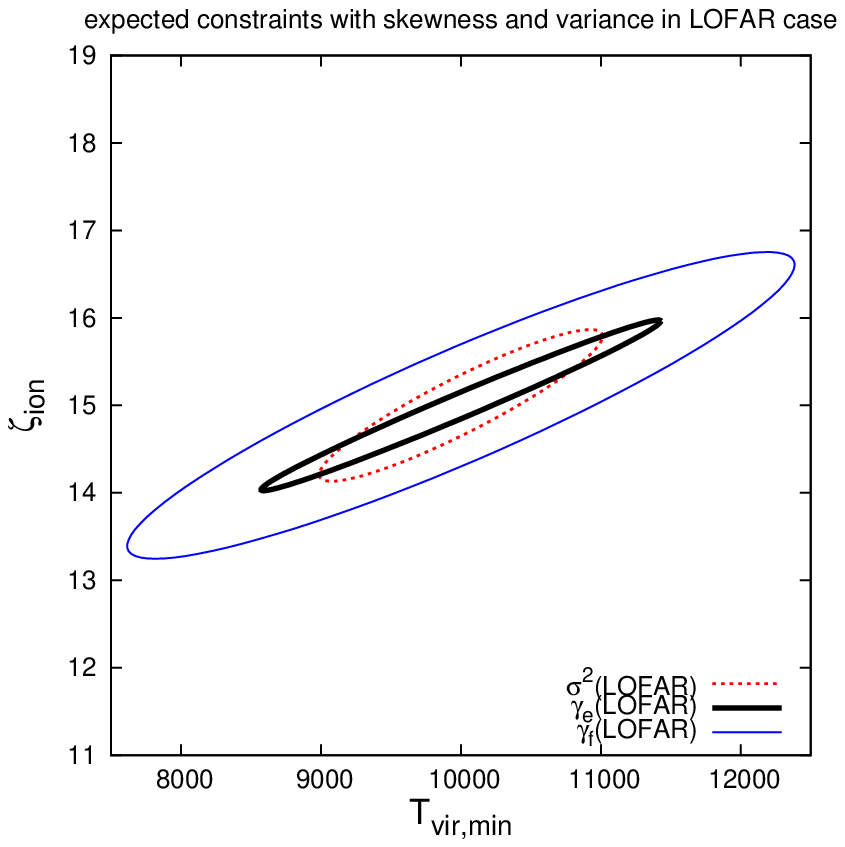}
\includegraphics[width=10cm]{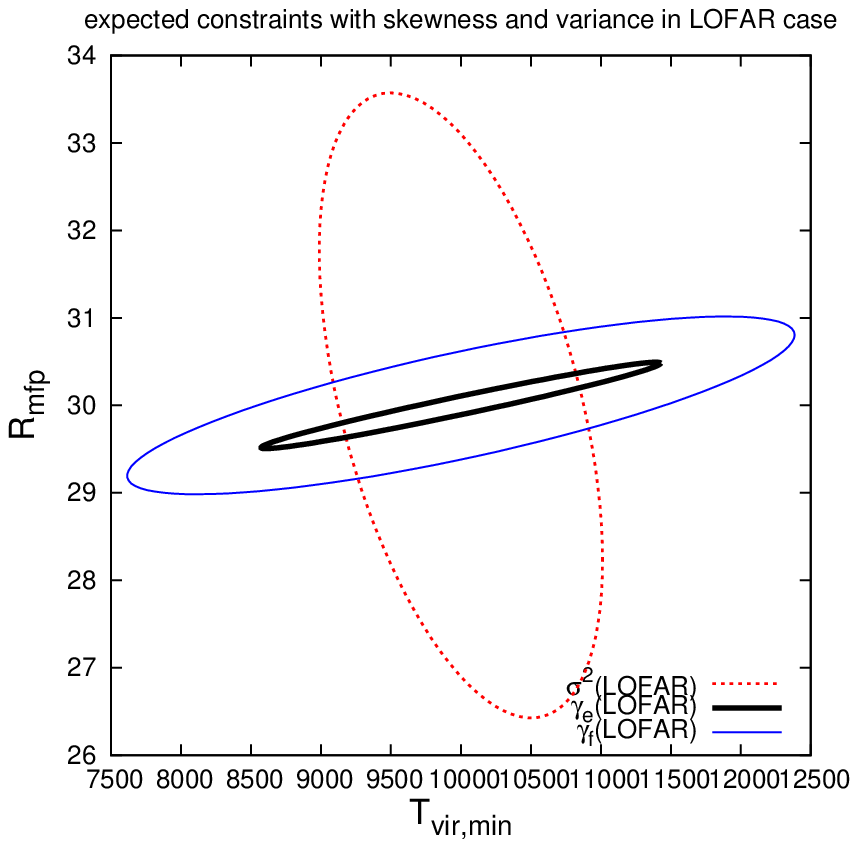}
\includegraphics[width=10cm]{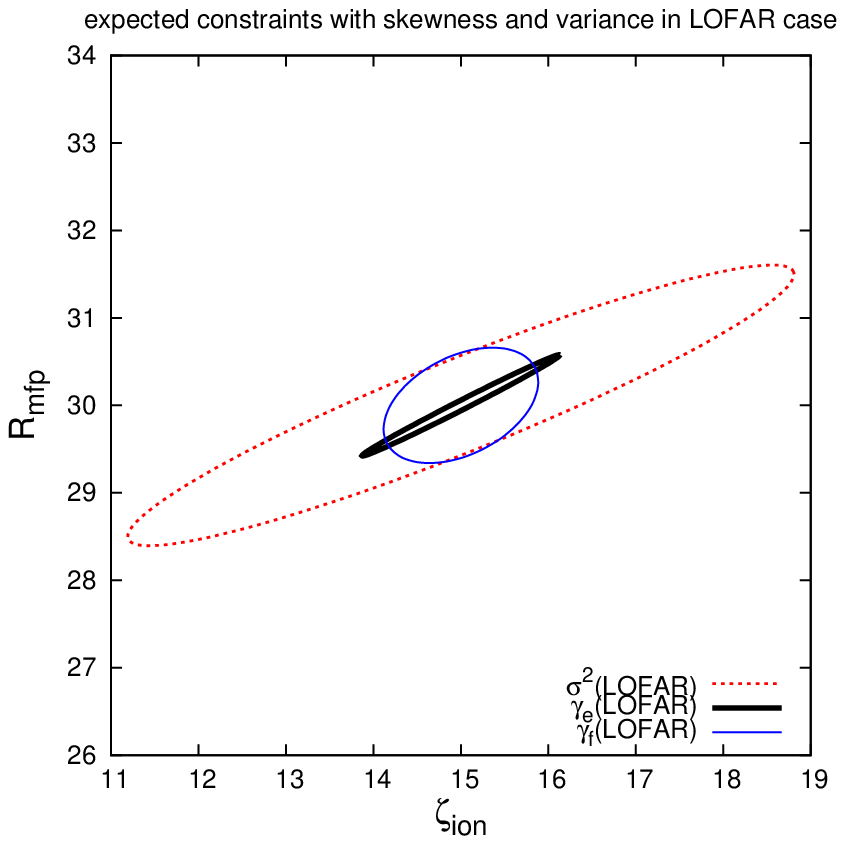}
\end{center}
\vspace{13mm}
\caption{Same as Fig. \ref{fig:MWA1} but for the LOFAR.}
\label{fig:LOFAR1}
\end{figure}

The LOFAR can put better constraints as can be seen in Fig. \ref{fig:LOFAR1}, while the qualitative features of the constraints are similar to those from the MWA. The combination of three observables leads to the parameter constraints of $T_{\rm vir,min} = 10^4 \pm 1000~{\rm K}$, $\zeta_{\rm ion} = 15.0 \pm 1.0$ and $R_{\rm mfp} = 30.0 \pm 0.5~{\rm Mpc}$. All the three parameters are constrained very well.

\begin{figure}
  \begin{center}
\includegraphics[width=10cm]{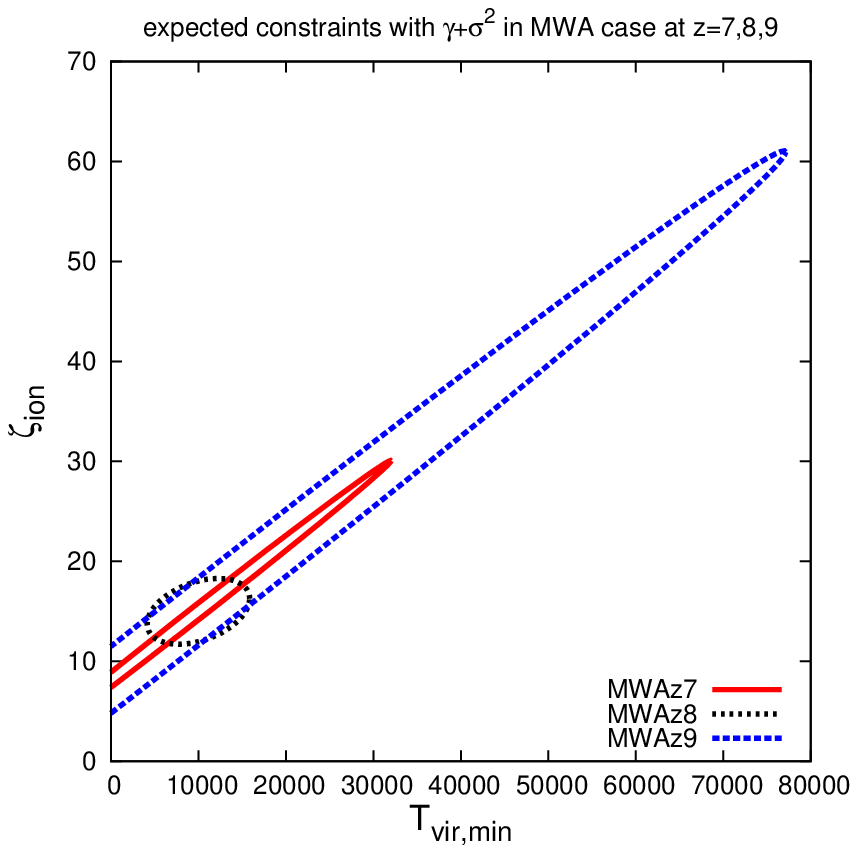}
\includegraphics[width=10cm]{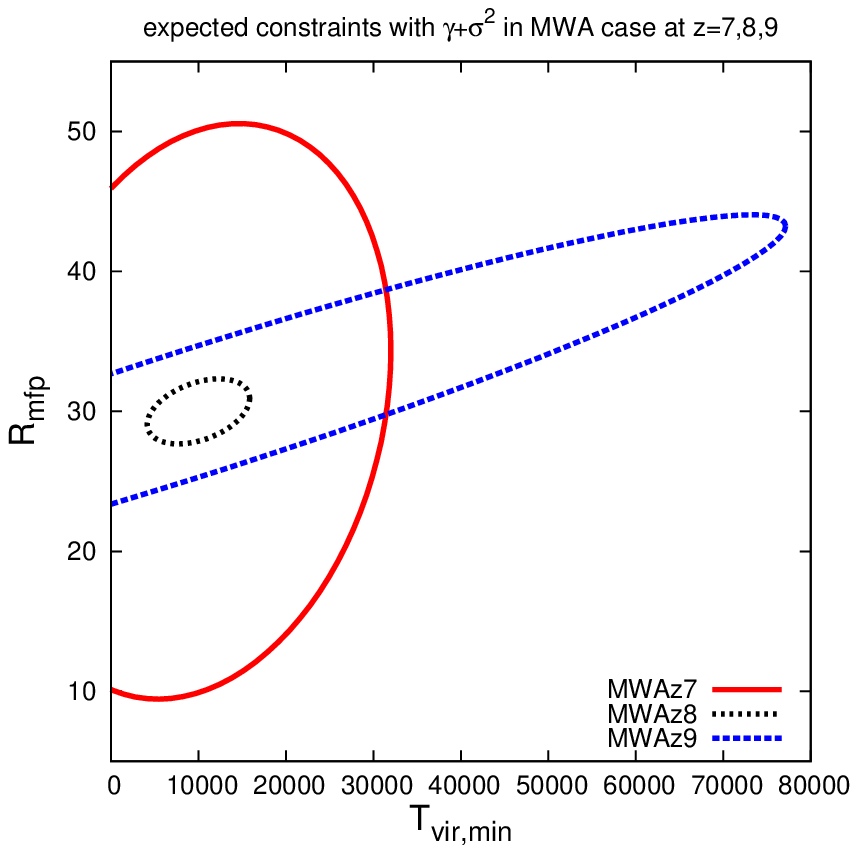}
\includegraphics[width=10cm]{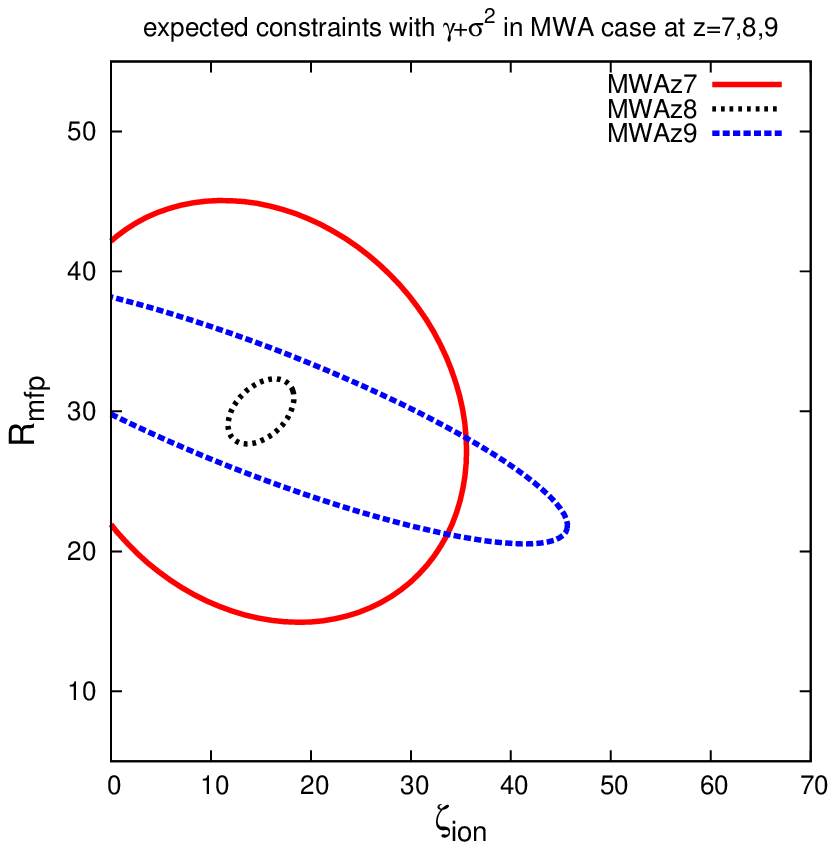}
\end{center}
\vspace{13mm}
\caption{Expected 1-$\sigma$ constraints on EoR parameters with the skewness and variance for 1000-hour observation of the MWA. The thick solid, dotted and dashed lines represents the constraints from $z = 7, 8$ and $9$, respectively.}
\label{fig:MWA2}
\end{figure}

\begin{figure}
\begin{center}
\includegraphics[width=10cm]{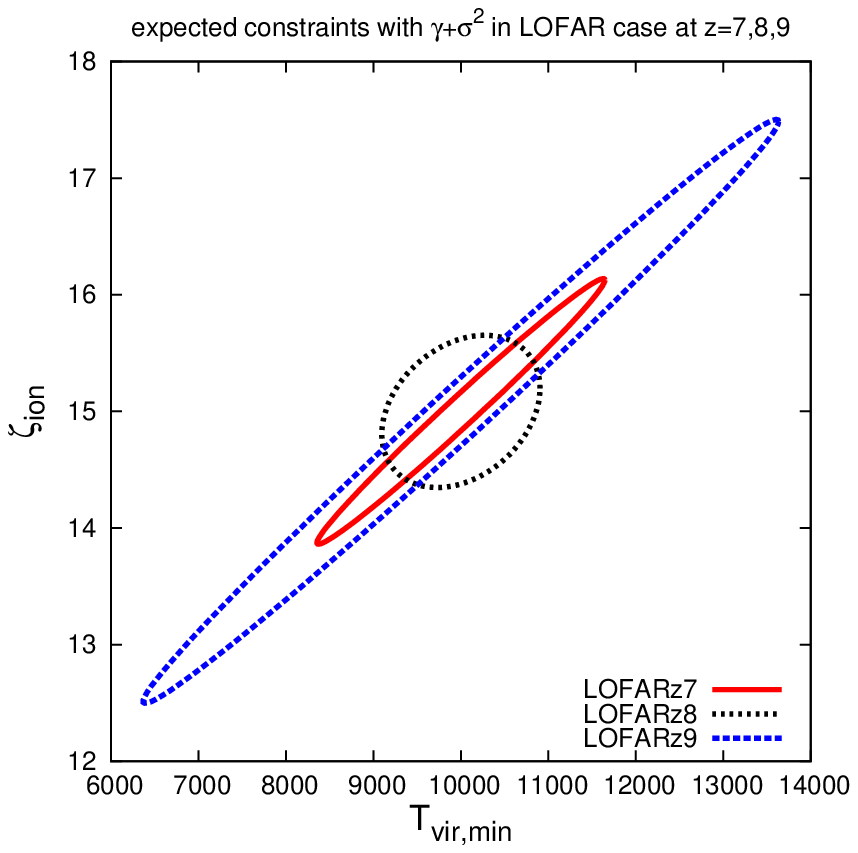}
\includegraphics[width=10cm]{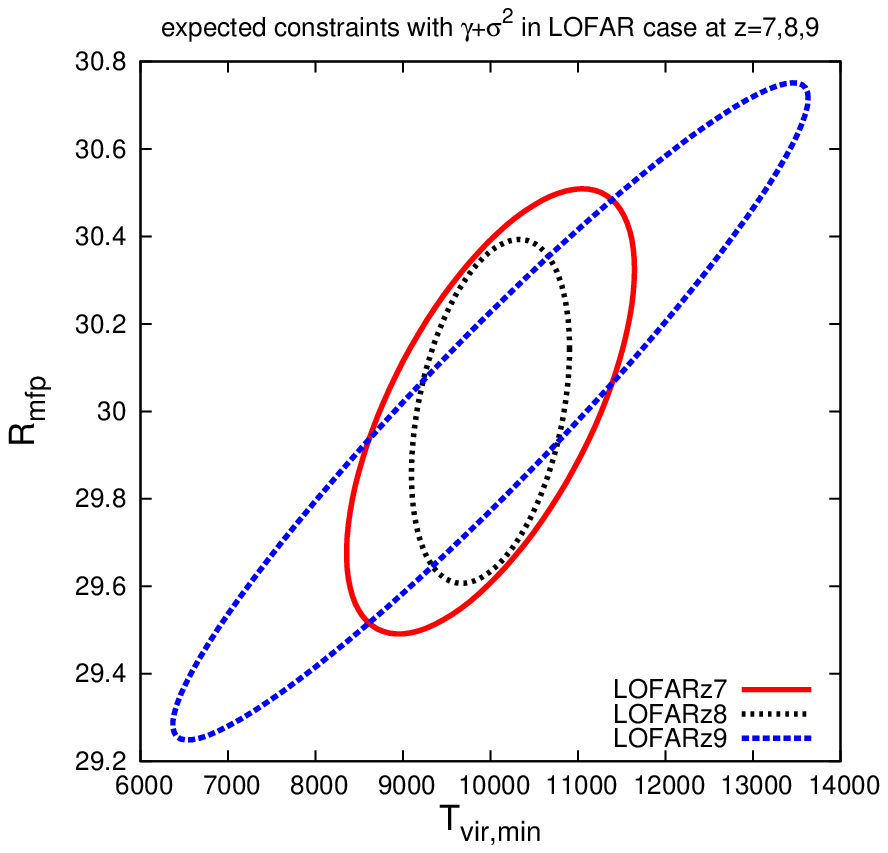}
\includegraphics[width=10cm]{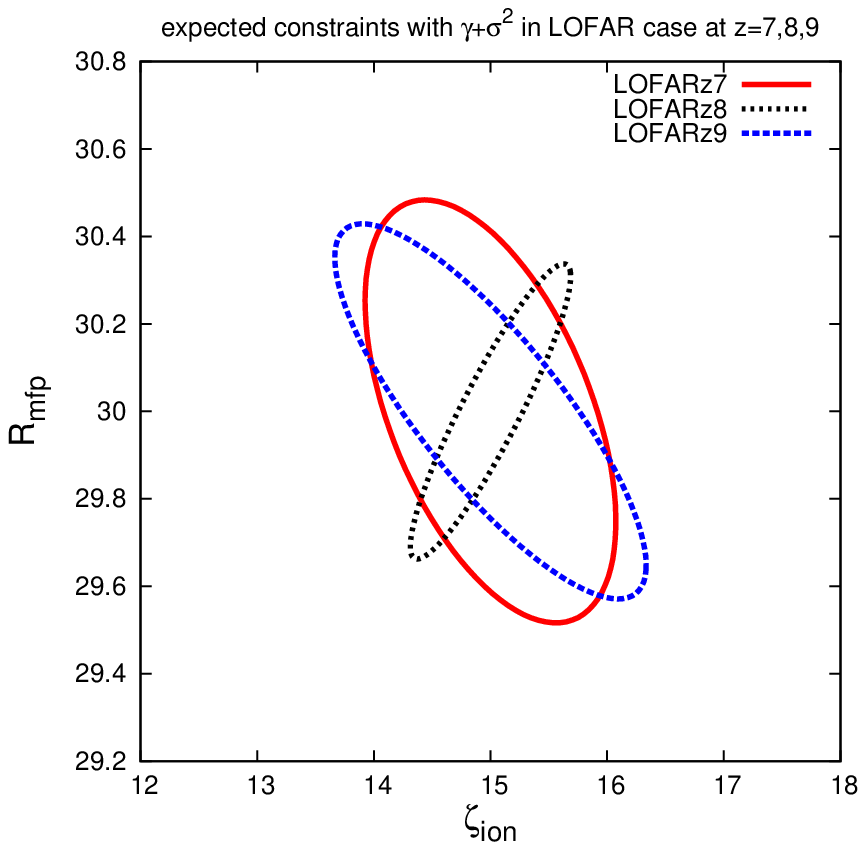}
\end{center}
\vspace{13mm}
\caption{Same as Fig. \ref{fig:MWA2} but for the LOFAR.}
\label{fig:LOFAR2}
\end{figure}

In order to obtain a deeper understanding of the constraints, we show the contribution of the observable quantities of each redshift in Figs. \ref{fig:MWA2} and \ref{fig:LOFAR2}. We can see that the constraints from $z=8$ are the tightest, which is reasonable considering the S/N ratios. While each redshift has a different degeneracy on the parameter space, the constraints from $z=8$ dominate over those from other redshifts. Thus, a combination of the three redshifts will not improve the constraints significantly compared with the case using only $z=8$ quantities, especially in the case of the MWA observation. Obviously, the peak redshift depends on the model parameters and we need observations of a broad range of the redshift.

\section{Summary and Discussion}

In this paper, we investigated the potential of the variance and skewness of the probability distribution function of the 21cm brightness temperature for constraining the model parameters of a public code 21cmFAST, which we used to obtain the brightness temperature maps. These statistical quantities are easy to calculate from the observed visibility and thus suitable for the early exploration of the EoR with ongoing telescopes such as the MWA and LOFAR. Evaluating both the signal and noise for the variance and skewness, we focused on a redshift range of $z=7-9$ where the S/N ratios are relatively high. We showed that a combination of the variance and skewness can strongly constrain the EoR model parameters such as the minimum virial temperature of halos which host luminous objects, ionizing efficiency and mean free path of ionizing photons in the IGM. In particular, the LOFAR can measure these parameters with errors less than $10\%$. For our fiducial values of the model parameters, the constraints from $z=8$ are dominant over the other redshifts.

Here we used a public code, 21cmFAST, which is based on a relatively simple reionization model. We will need large-scale full numerical simulations to interpret observational data in detail. However, full simulations need enormous calculations because of complex astrophysical effects such as the radiative feedback and recombination. Therefore, they are not currently suitable for the parameter search by fitting the observational data from ongoing telescopes. Although 21cmFAST is a simple model, the resulting 21cm signal is well consistent with that from more sophisticated simulations at large scales ($\gtrsim 1~{\rm Mpc}$) and it is believed to describe the essence of the physical processes related to the EoR. Therefore, the fitting of observational data with 21cmFAST is able to not only reduce the computational cost but also extract information on the key ingredients of the EoR physics, which will greatly help full numerical simulations.

In our analysis, the errors associated with the foreground subtraction and calibration were not considered and only the thermal noise was taken into account for Fisher analysis. This is because there has been no study on the foreground and other systematics associated with the bispectrum and skewness so that we need to take only the thermal noises in order to treat the variance and skewness evenly. Therefore, the parameter constraints are much stronger than those obtained from the power spectrum in the previous studies \citep{2014ApJ...782...66P,2015MNRAS.449.4246G}. For example, \citet{2014ApJ...782...66P} performed Fisher analysis considering future measurements of the power spectrum of the brightness temperature in order to predict constraints on the same parameter set as in this work. They showed that their "optimistic foreground model", which is almost free from the foreground contamination, results in parameter constraints which are better by a factor of 3 and 10 than their "moderate foreground model", which takes the signal only within the "EoR window" into account, for the MWA and LOFAR, respectively. Comparing their case with "optimistic foreground model" to our results, our constraints are weaker by a factor of 3 - 4 for the MWA and 1.2 - 1.4 for the LOFAR. This would be because the variance is an integrated quantity and has less information than the power spectrum.

The foreground and systematics are essential for more practical studies on the detectability of the variance and skewness and each telescope has its own strategy to overcome these obstacles. Obviously, parameter constraints from real observations are highly dependent not only on the telescope's sensitivity but on how we can reduce the systematic errors. Although the systematics concerned with the power spectrum and variance have been studied extensively, we need to study them also for the bispectrum and skewness.

\section*{Acknowledgement}
This work is supported by Grand-in-Aid from the Ministry of Education, Culture, Sports, and Science and Technology (MEXT) of Japan, No. 24340048, No. 26610048, No. 15H05896 (K.T.) and No. 25-3015 (H.S.).


\end{document}